\documentclass[12pt]{article}
\usepackage{graphicx}
\usepackage{wrapft}
\usepackage{wrapfig}
\usepackage{cite}
\usepackage{amsmath}
\usepackage{amssymb}
\usepackage{amsfonts}
\usepackage{physics}
\usepackage{ulem}
\usepackage{color}
\numberwithin{equation}{section}
\def\be{\begin{eqnarray}}
\def\ee{\end{eqnarray}}
\def\eq{\label}
\def\nn{\nonumber  \\}
\def\abstract#1{\vskip 7mm 
\begin{center}{\large Abstract}\par \bigskip
\begin{minipage}[c]{12cm}
\small #1
\end{minipage}
\end{center}
}
\def\title#1{\begin{center}{\Large\bf #1}\end{center}}
\def\author#1{\vskip 5mm \begin{center}{#1}\end{center}}
\def\address#1{\begin{center}{\it #1}\end{center}}
\newcommand{\bfr}{\begin{flushright}}
\newcommand{\efr}{\end{flushright}}
\begin{document}
%%%%%%%%%%%%%%%%%%%%%%%%%%%%%%%%%%%%%%%%%%%%%%%%%%%%%%%%%%%%%%%%%%%%%%%%%%%%%%%
%\title{Understanding of Soliton Equations from the Gauge Theoretical Viewpoint I}
\title{Gauge-Theoretical Method in Solving Zero-curvature Equations II \\[1ex]\large{---Non-Weyl Class Solutions of the Static Einstein-Maxwell Equations---}}
%\vspace{2.5cm}
\author{
T.~Azuma
\footnote{azuma@dokyo.ac.jp}
}
\address{Dokkyo University\\
1-1 Gakuencho, Soka, Saitama 340-0042, Japan 
}
\author{
T.~Koikawa
\footnote{koikawa@otsuma.ac.jp}
}
\address{Institute of Human Culture Studies, Otsuma Women's University\\
12 Sanban-cho, Chiyoda-ku, Tokyo 102-8357, Japan}
\hspace{0.54cm}
\vspace{4cm}
\abstract{
The gauge-theoretical method introduced in our previous paper is applied to solve the axisymmetric and static Einstein-Maxwell equations. We obtain the solutions of the non-Weyl class, where the gravitational and electric or magnetic potentials are not functionally related. In the electrostatic case, we show that the obtained solution coincides with the solution given by Bonnor in 1979. In the magnetostatic case, we present a solution describing the gravitational field created by two magnetically charged masses. In this solution, we present a case in which the Dirac string does not stretch to spatial infinity but lies between the magnetically charged masses.}
\thispagestyle{empty}
\newpage
%%%%%%%%%%%%%%%%%%%%%%%%%%%%%%%%%%%%%%%%%%%%%%%%%%%%%%%%%%%%%%%%%%%%%%%%%%%
\section{Introduction}
\renewcommand{\thefootnote}{\fnsymbol{footnote}}
In our previous paper\cite{AK}, we reviewed how the inverse scattering method was developed in the soliton equations and the Einstein equation, citing several foundational works\cite{miura,GGKM,ZakShab1,ZakShab2,Wadachi1,Wadachi2,akns,aks,BZ1,BZ2,BV}.
We also introduced a gauge-theoretical formulation to solve both soliton equations, such as the nonlinear Schr\"odinger equation,  and the Einstein equation.

We considered two linear differential equations for a matrix wave function
$\Psi=\Psi(\lambda,x_1,x_2)$, given by
\be
\hat D_i\Psi&=&\mathcal{A}_i\Psi,\quad (i=1,2)\label{ldeq}
\ee
where $\hat D_i (i=1,2)$ are a pair of commuting differential operators, which are linear combinations of $\partial /\partial x_1, \partial /\partial x_2$ and $\partial /\partial\lambda$, and $\mathcal A_i=\mathcal A_i(\lambda,x_1,x_2) (i=1,2)$ are matrix-valued gauge potentials. From the compatible condition of Eq.(\ref{ldeq})
\be
\left[\hat D_1-\mathcal A_1, \hat D_2-\mathcal A_2\right]\Psi=0,
\ee
we obtain the zero-curvature equation
\be
\hat D_1\mathcal A_2-\hat D_2\mathcal A_1+[\mathcal A_2, \mathcal A_1]=0.\label{zceq}
\ee
We demonstrated that the zero-curvature equation (\ref{zceq}) leads to the nonlinear Schr\"odinger equation and one of the equations comprising the vacuum Einstein equation with axial symmetry. 

We also considered  a singular gauge transformation that preserves the zero-curvature equation (\ref{zceq}):
\be
\Psi_0 &\to& \Psi=\mathcal{G}\Psi_0,\label{gtp}\\
\mathcal{A}_{0i} &\to& \mathcal{A}_i=\mathcal{G} \mathcal{A}_{0i}\mathcal{G}^{-1}-(\hat D_i\mathcal{G})\mathcal{G}^{-1},\quad(i=1,2)\label{gta}
\ee
where $\mathcal{G}=\mathcal{G}(\lambda,x_1,x_2)$ is a matrix-valued function.
We showed that this transformation generates a new wave function $\Psi$ from a trivial or known  wave function $\Psi_0$. 

As an application of this formulation, we solved the static and axisymmetric Einstein-Maxwell equations with magnetic charge. We assumed that the gravitational potential $f$ and the magnetic potential $\psi$ are functionally related, i.e., $f=f(\psi)$. The obtained solution belongs to the Weyl class and describes a magnetically charged monopole black hole with a Dirac string extending from the black hole to $-\infty$ or $+\infty$ along the symmetry axis (the region (i) or (iii) in Fig.1 in Ref.\cite{AK}) in the 2-soliton case. In the 4-soliton case, the solution describes the gravitational field generated by two magnetically charged masses and a strut between them along the symmetry axis. In this solution one of the Dirac strings lies between the masses, and the other extends from each mass to $-\infty$ or $+\infty$ (the regions (iii) and (i), or (iii) and (v) in Fig.2 in Ref.\cite{AK}), or both Dirac strings extend from each mass to $\pm\infty$ along the symmetry axis(the regions (i) and (v) in Fig.2 in Ref.\cite{AK}).

In this paper, we consider non-Weyl class solutions of the static axisymmetric Einstein-Maxwell equations, in which the gravitational and electric or magnetic potentials are not functionally related. In the electric case, it is known that the solution of the non-Weyl class is the solution presented by Bonnor in 1979\cite{B1}, which is distinct from the Reissner-Nordstr\" om solution\cite{reiss, nord}. This solution describes a gravitational field generated by two electrically charged masses and exhibits unusual properties not found in classical electrostatics\cite{BS, B2}. 

In the following section, we show that the gauge-theoretical formulation can be applied to solve the equations even in the non-Weyl class case, and that the 2-soliton solution in the electrostatic case reproduces Bonnor's solution. In section 3, in the magnetic case we present a new magnetostatic solution of the Einstein-Maxwell equations, which differs from the solutions given in our previous paper. The 2-soliton solution describes a gravitational field generated by two magnetically charged masses and the strut between them. We also show that there exists a solution in which the Dirac string does not extend to infinity but lies entirely between the two magnetically charged masses. In the final section, we provide a brief summary and discussion, focusing on the difference between the 2-soliton non-Weyl class and the 4-soliton Weyl class solution.
%%%%%%%%%%%%%%%%%%%%%%%%%%%%%%%%%%%%%%%%%%%%%%%%%%%%%%%%%%%%%%%%%%%%%%%%%%%
\section{Non-Weyl class solution with electric \\ charge}
%%%%%%%%%%%%%%%%%%%%%%%%%%%%%%%%%%%%%%%%%%%%%%%%%%%%%%%%%%%%%%%%%%%%%%%%%%%
In this section, we apply the gauge-theoretical method to solve the static and axisymmetric Einstein-Maxwell equations with electric charge. We do not impose a functional relationship between the gravitational and electric potentials. We demonstrate that the resulting solution is  one presented by Bonnor\cite{B1}. 

We begin with the source-free Einstein-Maxwell equations:
\be
&&R_{\mu\nu}=2\left( F_\mu{}^\alpha F_{\nu\alpha}
-\frac{1}{4}g_{\mu\nu}F^{\alpha\beta}F_{\alpha\beta}\right),\label{EMeq}\\
&&F^{\mu\nu}{}_{;\nu}=0,\\
&&F_{\mu\nu;\lambda}+F_{\nu\lambda;\mu}+F_{\lambda\mu;\nu}=0,\\
&&F_{\mu\nu}=A_{\nu;\mu}-A_{\mu;\nu},\label{Feq}
\ee
in the metric
\be
ds^2 =- fdt^2+f^{-1}[e^k(d\rho^2+dz^2) + \rho^2d\phi^2], \label{metric}
\ee
where $f$ and $k$ are functions of $\rho$ and $z$. 
In the electrostatic case, the equations reduce to
\be
&&(\ln f),_{\rho\rho} + \rho^{-1}(\ln f),_{\rho} + (\ln f),_{zz}
 = 2f^{-1}(\chi,_\rho^2+\chi,_z^2),\label{eqlnf1}\\
&&k,_\rho = {\rho \over 2}[(\ln f),_\rho^2-(\ln f),_z^2] 
 -2\rho f^{-1}(\chi,_\rho^2-\chi,_z^2), \label{eqkrho1}\\
&&k,_z = \rho(\ln f),_\rho(\ln f),_z -4\rho f^{-1}\chi,_\rho\chi,_z,\label{eqkz1} \\
&&\chi,_{\rho\rho} + \rho^{-1}\chi,_{\rho}+\chi,_{zz}
 =\chi,_{\rho}(\ln f),_{\rho} + \chi,_z(\ln f),_z,\label{eqchi}\\
 &&\chi,_{\rho z}=\chi,_{z\rho},
\ee
where $\chi=-A_t$.

If we assume $f=f(\chi)$, we obtain the Reissner-Nordstr\"om solution,  which belongs to the Weyl class\cite{AK0}. However we do not adopt this assumption here, aiming instead to derive a non-Weyl class solution. 

To apply the gauge-theoretical method to solve Eqs.(\ref{eqlnf1})-(\ref{eqchi}), we introduce a function $\tilde\chi(\rho,z)$ by
\be
\tilde\chi,_z=-\rho f^{-1}\chi,_\rho,\quad
\tilde\chi,_\rho=\rho f^{-1}\chi,_z,\label{tildechi}
\ee
and rewrite Eqs.(\ref{eqlnf1})-(\ref{eqchi}) as
\be
&&(\ln f),_{\rho\rho} + \rho^{-1}(\ln f),_{\rho} + (\ln f),_{zz}
 = 2\rho^{-2}f(\tilde\chi,_\rho^2+\tilde\chi,_z^2),\label{eqlnf2}\\
&&k,_\rho = {\rho \over 2}[(\ln f),_\rho^2-(\ln f),_z^2] 
 +2\rho^{-1}f(\tilde\chi,_\rho^2-\tilde\chi,_z^2), \label{eqkrho2}\\
&&k,_z = \rho(\ln f),_\rho(\ln f),_z +4\rho^{-1}f\tilde\chi,_\rho\tilde\chi,_z,\label{eqkz2} \\
&&\tilde\chi,_{\rho\rho} - \rho^{-1}\tilde\chi,_{\rho}+\tilde\chi,_{zz}
 =-\tilde\chi,_{\rho}(\ln f),_{\rho} - \tilde\chi,_z(\ln f),_z.\label{eqchi2}
\ee
We further introduce $2\times 2$ matrices $h(\rho,z)$, $U(\rho,z)$ and $V(\rho,z)$ as follows
\be
&&h=\mqty(h_{00} & h_{01} \\ h_{10} & h_{11})=\mqty(f^{1/2} & f^{1/2}\tilde\chi \\ f^{1/2}\tilde\chi & f^{-1/2}\rho^2+f^{1/2}\tilde\chi^2),\label{matrixh}\\
&&U=\rho h,_\rho h^{-1},\qquad V=\rho h,_zh^{-1}.\label{UVeq}
\ee
We then find that Eqs.(\ref{eqlnf2})-(\ref{eqchi2}) are reduced to
\be
&&(\rho h,_\rho h^{-1}),_\rho+(\rho h,_zh^{-1}),_z=0,\label{eqh1}\\
&&k,_\rho=4(\ln h_{00}),_\rho-\frac{4}{\rho}+\frac{1}{\rho}\Tr(U^2-V^2),\label{eqkrho1}\\
&&k,_z=4(\ln h_{00}),_z+\frac{2}{\rho}\Tr(UV).\label{eqkz1}
\ee

These definitions allow us to rewrite the Einstein-Maxwell equations in terms of matrix equations derived from the zero-curvature condition.
We then define the differential operators and gauge potentials in Eq.(\ref{ldeq}) with $x_1=\rho$ and $x_2=z$ by
\be
&&\hat D_1=\partial_z-\frac{2\lambda^2}{\lambda^2+\rho^2}\partial_\lambda,\quad
\hat D_2=\partial_\rho+\frac{2\lambda\rho}{\lambda^2+\rho^2}\partial_\lambda,\label{gp1}\\
&&\mathcal{A}_1=\frac{\rho V-\lambda U}{\lambda^2+\rho^2},\quad\quad\quad \
\mathcal{A}_2=\frac{\rho U+\lambda V}{\lambda^2+\rho^2}.\label{gp2}
\ee
Requiring that the zero-curvature condition (\ref{zceq}) holds for arbitrary $\lambda$, we obtain
\be
\left(\frac{U}{\rho}\right),_z-\left(\frac{V}{\rho}\right),_\rho+\left[\left(\frac{U}{\rho}\right),\label{zcuv}
\left(\frac{V}{\rho}\right)\right]&=&0,\\
U,_\rho+V,_z&=&0.\label{equv}
\ee
When $\lambda=0$, we have $\hat D_1=\partial_z, \hat D_2=\partial_\rho, \mathcal{A}_1=V/\rho$ and $\mathcal{A}_2=U/\rho$, and then Eq.(\ref{ldeq}) reads
\be
\partial_z\Psi(0,\rho,z)=\frac{V}{\rho}\Psi(0,\rho,z),\quad \partial_\rho\Psi(0,\rho,z)=\frac{U}{\rho}\Psi(0,\rho,z).\label{lmd0}
\ee
By comparing Eq.(\ref{lmd0}) with Eq.(\ref{UVeq}) we find that
\be
h(\rho,z)=\Psi(0,\rho,z).
\ee

We now consider the gauge transformation (\ref{gtp}) to obtain a new soliton solution $h(\rho,z)$ from a certain trivial solution $h_0(\rho,z)$. The wave function $\Psi_0(\lambda,\rho,z)$ corresponding to $h_0(\rho,z)$ satisfies the relations
\be
h_0(\rho,z)=\Psi_0(0,\rho,z),
\ee
and
\be
\hat D_i\Psi_0=\mathcal{A}_{i0}\Psi_0, \quad (i=1,2)\label{ldeq0}
\ee
where $\mathcal{A}_{0i} (i=1,2)$ are given by
\be
\mathcal{A}_{01}=\frac{\rho V_0-\lambda U_0}{\lambda^2+\rho^2},\quad\quad
\mathcal{A}_{02}=\frac{\rho U_0+\lambda V_0}{\lambda^2+\rho^2},\label{A0}
\ee
with
\be
\ U_0=\rho h_0,_\rho h_0^{-1},\quad\quad\quad V_0=\rho h_0,_zh_0^{-1}.\label{U0V0}
\ee
Substituting Eq.(\ref{gtp}) into Eq.(\ref{ldeq}) and using the relation (\ref{ldeq0}),
we have
\be
\hat D_i\mathcal{G}=\mathcal{A}_i\mathcal{G}-\mathcal{G}\mathcal{A}_{0i}.\quad (i=1,2)\label{ldeq2}
\ee
In order to obtain the soliton solutions with soliton number $N$, we introduce the singular gauge transformation of the form
\be
\mathcal{G}(\lambda,\rho,z)=I+\sum_{k=1}^{N}\frac{\mathcal{R}_k(\rho,z)}{\lambda-\mu_k(\rho,z)}\eq{Npgt},
\ee
where $\mu_k(\rho,z)$ are the poles in $\lambda$ space and $\mathcal{R}_k(\rho,z)$ are the residues. Substituting Eq.(\ref{Npgt}) into Eq.(\ref{ldeq2}) under the condition that $h(\rho,z)$ is a symmetric matrix yields expressions for $\mu_k(\rho,z)$ and $\mathcal{R}_k(\rho,z)$. The detailed procedure described below is provided in Refs.\cite{BZ2} and \cite{BV}. The poles 
$\mu_k(\rho,z) (k=1,2,\cdots,N)$ are solutions of a pair of differential equations
\be
\mu_k,_\rho=\frac{2\rho\mu_k}{\mu_k^2+\rho^2},\quad\mu_k,_z=-\frac{2\mu_k^2}{\mu_k^2+\rho^2},\label{murhoz}
\ee
and satisfy the relation
\be
\mu_k^2-2(w_k-z)\mu_k-\rho^2=0,\label{eqmuk}
\ee
where $w_k$ are arbitrary constants.
The $2\times 2$ matrices $\mathcal{R}_k(\rho,z) (k=1,2,\cdots,N)$ are written as
\be
(\mathcal{R}_k)_{ab}=(\mathcal{N}^{(k)})_a(\mathcal{M}^{(k)})_b.\quad (a,b=0,1)\label{Rk}
\ee
In Eq.(\ref{Rk}), the two-component vectors $\mathcal{M}^{(k)}$ are given by
\be
\mathcal{M}^{(k)}=\Psi_0^{-1}(\mu_k,\rho,z)\mathcal{P}^{(k)},
\ee
where $\Psi_0^{-1}$ denotes the matrix inverse to $\Psi_0$ and 
$\mathcal{P}^{(k)}$ are two-component vectors $\mathcal{P}^{(k)}$ whose components 
$P^{(k)}_0$ and $P^{(k)}_1$ are arbitrary constants. $\mathcal{N}^{(k)}$ in Eq.(\ref{Rk}) are also two-component vectors given by
\be
\mathcal{N}^{(k)}=\sum_l^N(\Gamma^{-1})_{lk}\mu_l^{-1}h_0(\rho,z)\mathcal{M}^{(l)},
\ee
where $\Gamma^{-1}$ is an inverse to $N\times N$ matrix $\Gamma$ defined by
\be
(\Gamma)_{kl}=\frac{\tilde{\mathcal{M}}^{(k)}h_0(\rho,z)\mathcal{M}^{(l)}}
{\rho^2+\mu_k\mu_l}.
\ee
Here $\tilde{\mathcal{M}}^{(k)}$ denotes a transpose to $\mathcal{M}^{(k)}$. Then we obtain $\Psi$ that corresponds to a new soliton solution $h(\rho,z)$ with soliton number $N$ from $\Psi_0$ that corresponds to a certain trivial solution $h_0(\rho,z)$. The new solution is given by
\be
h(\rho,z)=\mathcal{G}(0,\rho,z)\Psi_0(0,\rho,z)=\left(I-\sum_k^N\frac{\mathcal{R}_k}{\mu_k}\right)h_0(\rho,z),\label{Nsh}
\ee
and its components are
\be
h_{ab}=(h_0)_{ab}-\sum_{k,l}^N(h_0\mathcal{M}^{(l)})_a\mu_l^{-1}(\Gamma^{-1})_{lk}(h_0\mathcal{M}^{(k)})_b\mu_k^{-1}.\quad (a,b=0,1)\label{Nshc}
\ee

We now study the soliton solutions obtained by using the gauge-theoretical method mentioned above. We adopt $h_0=\mathrm{diag}(1,\rho^2)$ as a trivial solution. 
Substituting this $h_0$ into Eqs.(\ref{A0}) and (\ref{U0V0}) we solve the differential equation (\ref{ldeq0}):
\be
\Psi_0=\mqty(1 & 0 \\ 0 & \rho^2-2z\lambda-\lambda^2),
\ee
and using the relation (\ref{eqmuk}) we have
\be
\mathcal{M}^k=\mqty(q_k \\ -p_k\mu_k^{-1}), 
\ee
and
\be
\Gamma_{kl}=\frac{p_kp_l\rho^2+q_kq_l\mu_k\mu_l}{\mu_k\mu_l(\rho^2+\mu_k\mu_l)},
\ee
where we have set the constants $P_0^{(k)}$ and $P_1^{(k)}$ as $P_0^{(k)}=q_k$ and $P_1^{(k)}=2w_kp_k$. Since the obtained solution (\ref{Nshc}) does not always satisfy the condition 
$\det h=\rho^2$, we normalize $h(\rho,z)$ by multiplying it by $\rho(\det h)^{-1/2}$.

When $N=1$ in Eq.(\ref{Nshc}), we have the 1-soliton solution for $h(\rho,z)$. However, we find that the solutions for $f$ and $k$ do not satisfy the asymptotic flatness condition. When $N=2$, we  obtain the 2-soliton solution for $h(\rho,z)$ given by
\be
h_{00}&=&\frac{\mu_1\mu_2[\rho^2(\mu_2-\mu_1)^2(p_1p_2-q_1q_2)^2+(\rho^2+\mu_1\mu_2)^2(p_1q_2-p_2q_1)^2]}
{(\mu_2-\mu_1)^2(p_1p_2\rho^2+q_1q_2\mu_1\mu_2)^2-(\rho^2+\mu_1\mu_2)^2(p_1q_2\mu_2-p_2q_1\mu_1)^2},\label{h00}
\nn\\
h_{01}&=&-(\mu_2-\mu_1)(\rho^2+\mu_1\mu_2) \nonumber \\
&\times& \frac{p_2q_2(\mu_2^2+\rho^2)(p_1^2\rho^2+q_1^2\mu_1^2)-p_1q_1(\mu_1^2+\rho^2)(p_2^2\rho^2+q_2^2\mu_2^2)}
{(\mu_2-\mu_1)^2(p_1p_2\rho^2+q_1q_2\mu_1\mu_2)^2-(\rho^2+\mu_1\mu_2)^2(p_1q_2\mu_2-p_2q_1\mu_1)^2},\label{h01}
\nn
\ee
where
\be
\left\{
\begin{array}{l}
\mu_1=w_1-z+\sqrt{(w_1-z)^2+\rho^2}\\
\mu_2=w_2-z-\sqrt{(w_2-z)^2+\rho^2}.\label{mu12}
\end{array}
\right.
\ee
From Eq.(\ref{matrixh}) we can express $f$ and $\tilde\chi$ in terms of the components of the matrix $h$ as
\be
f=h_{00}^2,\quad \tilde\chi=h_{01}/h_{00}.\label{fchi}
\ee
By substituting Eqs.(\ref{h00}) and (\ref{h01}) into Eq.(\ref{fchi}), we obtain
\be
f&=&\left[
\frac{\mu_1\mu_2[\rho^2(\mu_2-\mu_1)^2(p_1p_2-q_1q_2)^2+(\rho^2+\mu_1\mu_2)^2(p_1q_2-p_2q_1)^2]}
{(\mu_2-\mu_1)^2(p_1p_2\rho^2+q_1q_2\mu_1\mu_2)^2-(\rho^2+\mu_1\mu_2)^2(p_1q_2\mu_2-p_2q_1\mu_1)^2}
\right]^2,\label{solf}\nn\\
\tilde\chi&=&-(\mu_2-\mu_1)(\rho^2+\mu_1\mu_2) \nn
&\times&
\frac{p_2q_2(\mu_2^2+\rho^2)(p_1^2\rho^2+q_1^2\mu_1^2)-p_1q_1(\mu_1^2+\rho^2)(p_2^2\rho^2+q_2^2\mu_2^2)}
{\mu_1\mu_2[\rho^2(\mu_2-\mu_1)^2(p_1p_2-q_1q_2)^2+(\rho^2+\mu_1\mu_2)^2(p_1q_2-p_2q_1)^2]},\label{solpsi}\nn
\ee
and the integrations of Eqs.(\ref{tildechi}), (\ref{eqkrho1}) and (\ref{eqkz1}) give
\be
\chi&=&(\mu_2-\mu_1)(\rho^2+\mu_1\mu_2)\nonumber \\
&\times&
\frac{p_2q_2(p_1^2-q_1^2)\mu_1(\mu_2^2+\rho^2)-p_1q_1(p_2^2-q_2^2)\mu_2(\mu_1^2+\rho^2)}
{(\mu_2-\mu_1)^2(p_1p_2\rho^2+q_1q_2\mu_1\mu_2)^2-(\rho^2+\mu_1\mu_2)^2(p_1q_2\mu_2-p_2q_1\mu_1)^2},\label{solchi}
\nn\\
e^k&=&K_2\left[\frac{(p_1p_2-q_1q_2)^2\rho^2(\mu_2-\mu_1)^2+(p_1q_2-p_2q_1)^2(\rho^2+\mu_1\mu_2)^2}
{(\mu_1^2+\rho^2)(\mu_2^2+\rho^2)}\right]^4,\label{solk}\nn
\ee
where $K_2$ is a constant.

If we introduce prolate spheroidal coordinates $(x,y)$ defined by
\be
\left\{
\begin{array}{l}
\rho=\sigma\sqrt{(x^2-1)(1-y^2)}\\
z-z_0=\sigma xy,\label{xy}
\end{array}
\right.\label{psc}
\ee
with
\be
\sigma=\frac{w_2-w_1}{2},\quad z_0=\frac{w_2+w_1}{2},\label{}
\ee
we then find that the solutions (\ref{solf}), (\ref{solchi}) and (\ref{solk}) are written as
\be
f&=&\left[\frac{c_2^2(x^2-1)+c_4^2(1-y^2)}{(c_2x+c_1)^2-(c_3+c_4y)^2}\right]^2,\label{fxy}\\
\chi&=&\frac{2(c_2c_3x-c_1c_4y)}{(c_2x+c_1)^2-(c_3+c_4y)^2},\label{chixy}\\
e^k&=&K_2\left[\frac{c_2^2(x^2-1)+c_4^2(1-y^2)}{x^2-y^2}\right]^4,\label{kxy}
\ee
where the constants $c_1$, $c_2$, $c_3$ and $c_4$ are given by
\be
\left\{
\begin{array}{l}
c_1=p_1p_2+q_1q_2\\
c_2=p_1p_2-q_1q_2\\
c_3=-(p_1q_2+p_2q_1)\\
c_4=-(p_1q_2-p_2q_1).\label{ceqs}
\end{array}
\right.
\ee
In Eq.(\ref{ceqs}), we note that the following relation holds:
\be
c_1^2+c_4^2=c_2^2+c_3^2.\label{crel}
\ee
If we define the quantities $A$, $B$ and $C$ introduced in the original paper by Bonnor\cite{B1} by
\be
A=\frac{c_4}{c_2},\quad B=\frac{c_3-c_1}{c_2},\quad C=\frac{c_3+c_1}{c_2},\label{}
\ee
and set $K_2=c_2^{-8}$, we then find that the solution is Bonnor's solution.

Bonnor's solution was known as a non-Weyl class solution of the static and axisymmetric Einstein-Maxwell equations and interpreted as describing the gravitational and electric fields created by a mass with an electric charge and dipole moment. We can see this by the asymptotic behaviors of the solutions at spatial infinity $\sqrt{\rho^2+z^2}\to\infty$:
\be
f&\sim& 1-\frac{2m}{\sqrt{\rho^2+z^2}}+\frac{2m^2+e^2}{\rho^2+z^2}-\frac{2l ez}{(\rho^2+z^2)^{3/2}},\label{finf}\\
\chi&\sim&\frac{e}{\sqrt{\rho^2+z^2}}-\frac{me}{\rho^2+z^2}+\frac{mlz}{(\rho^2+z^2)^{3/2}},
\label{chiinf}\\
e^k&\sim&1-\frac{(m^2-e^2)\rho^2}{(\rho^2+z^2)^2},\label{Qinf}
\ee
where $m$, $e$ and $l$ are constants given by
\be
m=2\frac{c_1}{c_2}\sigma,\quad
e=2\frac{c_3}{c_2}\sigma,\quad
l=-\frac{c_4}{c_2}\sigma,\label{Bmel}
\ee
and we have set $z_0=0$ in these behaviors. Since Bonnor presented this solution, it has been reexamined by other workers\cite{JC, CMR} and they revealed that the solution describes the fields by two charged masses. The individual masses and electric charges are defined, and the spacetime structure and some strange properties in the electric field are discussed in Refs.\cite {BS, B2}.

%%%%%%%%%%%%%%%%%%%%%%%%%%%%%%%%%%%%%%%%%%%%%%%%%%%%%%%%%%%%%%%%%%%%%
\section{Non-Weyl class solution with magnetic \\ charge}
%%%%%%%%%%%%%%%%%%%%%%%%%%%%%%%%%%%%%%%%%%%%%%%%%%%%%%%%%%%%%%%%%%%%%
In our previous paper, we presented the 2- and 4-soliton solutions of the Weyl class for the static and axisymmetric Einstein-Maxwell equations with magnetic charges using the gauge-theoretical method. We analyzed their spacetime structures and discussed the locations of the Dirac strings. In this section, we present a non-Weyl class solution of the Einstein-Maxwell equations with magnetic charge and examine the spacetime structure described by this solution.

The magnetostatic Einstein-Maxwell equations in the metric (\ref{metric}) are given by
\be
&&(\ln f),_{\rho\rho} + \rho^{-1}(\ln f),_{\rho} + (\ln f),_{zz}
 = 2 \rho^{-2}f(\psi,_\rho^2+\psi,_z^2),\label{eqlnf3}\\
&&k,_\rho = {\rho \over 2}[(\ln f),_\rho^2-(\ln f),_z^2] 
 +2\rho^{-1}f(\psi,_\rho^2-\psi,_z^2), \label{eqkrho3}\\
&&k,_z = \rho(\ln f),_\rho(\ln f),_z +4\rho^{-1}f\psi,_\rho\psi,_z,\label{eqkz3} \\
&&\psi,_{\rho\rho} - \rho^{-1}\psi,_{\rho}+\psi,_{zz}
 =-\psi,_{\rho}(\ln f),_{\rho} - \psi,_z(\ln f),_z],\label{eqpsi}\\
&&\psi,_{\rho z}=\psi,_{z\rho},\label{eqpsirhoz}
\ee
where we have set $A_t=A_\rho=A_z=0$ and written $A_\phi=\psi$. 
Note that these magnetostatic equations become identical to the electrostatic equations (\ref{eqlnf2})-(\ref{eqchi2}) when we identify $\tilde\chi$ with $\psi$. Therefore, we reuse the electrostatic solution $f$ from Eq.(\ref{solf}) and $k$ from Eq.(\ref{solk}) in the magnetostatic case. For the solution of the magnetic potential $\psi$, we have
\be
\psi=\tilde\chi+\kappa_m,\label{kappam}
\ee
where $\kappa_m$ is an arbitrary constant.

We now analyze the spacetime structure described by these solutions. We assume that $w_1=-\sigma$ and $w_2=\sigma$ in $\mu_1$ and $\mu_2$, and that $K_2=(p_1p_2-q_1q_2)^{-8}$ as in the electrostatic case. We also assume that $\kappa_m=0$ in Eq.(\ref{kappam}).  
At spatial infinity $r=\sqrt{\rho^2+z^2}\to\infty$, the solutions behave as
\be
f&\sim& 1-\frac{2m}{r}+\frac{2m^2+Q_m^2}{r^2}+\frac{2lQ_mz}{r^3},\label{fminf}\\
e^k&\sim&1-\frac{(m^2-Q_m^2)\rho^2}{r^4},\label{ekinf}\\
\psi&\sim&-2l-\frac{Q_mz}{r}-\frac{lm\rho^2}{r^3},\label{psiinf}
\ee
The magnetic field component behaves as
\be
B^r\sim\frac{Q_m}{r^2}-\frac{2mQ_m}{r^3}-\frac{2lmz}{r^4}.\label{Brinf}
\ee
The constants introduced in Eqs.(\ref{fminf})-(\ref{Brinf}) are given by
\be
m&=&2\frac{p_1p_2+q_1q_2}{p_1p_2-q_1q_2}\sigma,\label{cm}\\
Q_m&=&2\frac{p_1q_2+p_2q_1}{p_1p_2-q_1q_2}\sigma,\label{cQm}\\
l&=&\frac{p_1q_2-p_2q_1}{p_1p_2-q_1q_2}\sigma.\label{cl}
\ee
We note that in Eqs.(\ref{cm})-(\ref{cl}) the following relation holds:
\be
m^2-Q_m^2=4(\sigma^2-l^2).\label{mesl}
\ee
Introducing new variables $r_1=\sqrt{\rho^2+(z+\sigma)^2}$ and $r_2=\sqrt{\rho^2+(z-\sigma)^2}$, we find the behaviors at $r_1, r_2\to\infty$:
\be
f&\sim&-\left(m+\frac{lQ_m}{\sigma}\right)\frac{1}{r_1}
-\left(m-\frac{lQ_m}{\sigma}\right)\frac{1}{r_2},\label{finf2}\\
B^r&\sim&\left(Q_m+\frac{lm}{\sigma}\right)\frac{1}{2r_1^2}
+\left(Q_m-\frac{lm}{\sigma}\right)\frac{1}{2r_2^2}.\label{Brinf2}
\ee

We next consider the behaviors around the symmetry axis defined by $\rho=0$. In the following investigation, we assume that $\sigma>0$ and $l>0$ without loss of generality.
Dividing the axis into three regions, we have the following behaviors:

\noindent
i) $z<-\sigma$ region:
\be
f&=&\frac{16(\sigma^2-z^2)^2}{[(Q_m-2l)^2-(m-2z)^2]^2},\label{}\\
e^k&=&1,\label{}\\
\psi&=&-2l+Q_m\label{psi0d}.
\ee
ii) $-\sigma<z<\sigma$ region:
\be
f&=&\frac{16l^4(\sigma^2-z^2)^2}{(m^2\sigma^2-Q_m^2\sigma^2+4m\sigma^3+4\sigma^4-4Q_ml\sigma z-4l^2 z^2)^2},\label{}\\
e^k&=&\frac{l^8}{\sigma^8},\label{ek0m}\\
\psi&=&-\frac{\sigma(m+2\sigma)}{l}\label{psi0m}.
\ee
iii) $z>\sigma$ region:
\be
f&=&\frac{16(\sigma^2-z^2)^2}{[(Q_m+2l)^2-(m+2z)^2]^2},\label{uaxisf}\\
e^k&=&1,\label{}\\
\psi&=&-2l-Q_m\label{psi0u}.
\ee

These behaviors indicate that the solutions describe the gravitational field generated by two bodies with mass and magnetic charge located at $(\rho,z)=(0,-\sigma)$ and $(0,+\sigma)$. The total mass and magnetic charge are  $m$ and $Q_m$, respectively. The mass $m_1$ and magnetic charge $Q_{m1}$ of the lower body at $z=-\sigma$ are
\be
m_1&=&\frac{m}{2}+\frac{lQ_m}{2\sigma},\label{m1}\\
Q_{m1}&=&\frac{Q_m}{2}+\frac{lm}{2\sigma}.\label{Qm1}
\ee
The mass $m_2$ and magnetic charge $Q_{m2}$ of the upper body at $z=+\sigma$ are
\be
m_2&=&\frac{m}{2}-\frac{lQ_m}{2\sigma},\label{m2}\\
Q_{m2}&=&\frac{Q_m}{2}-\frac{lm}{2\sigma}.\label{Qm2}
\ee
The behaviors of $\psi$ in Eqs. (\ref{psi0d}), (\ref{psi0m}) and (\ref{psi0u}) show that the Dirac string spans the entire $z$-axis.
The behavior of $e^k$ in Eq.(\ref{ek0m}) indicates the presence of a strut between the masses: 
$-\sigma<z<\sigma$ on the axis unless $\sigma=l$. The spacetime structure described by the solutions is the same as that of Bonnor's solution except for the sign of charge. The metric function $f$ vanishes at the points defined by $r_1=0$ and $r_2=0$, indicating event horizons.
The curvature invariant $R^{\alpha\beta\gamma\delta}R_{\alpha\beta\gamma\delta}$ in the prolate spheroidal coordinates $(x,y)$ defined in Eq.(\ref{psc}) with $z_0=0$
\be
R^{\alpha\beta\gamma\delta}R_{\alpha\beta\gamma\delta}&=&
\frac{4096\sigma^{12}(x^2-y^2)^5}
{[(m+2\sigma x)^2-(Q_m+2ly)^2]^8[\sigma^2(x^2-1)+l^2(1-y^2)]^8}\nonumber\\
&\times&\Bigl({\rm{ polynomial \ of} } \ x,y \Bigl).\label{} 
\ee
shows no curvature singularity outside the horizon specified by $x>1$ if $\sigma>l$ and $m\pm Q_m>-2(\sigma-l)$.

Finally, we point out some interesting features of the solutions. In the case $\kappa_m=0$ in Eq.(\ref{kappam}), the Dirac string stretches from $-\infty$ to $+\infty$ as is seen in Eqs.(\ref{psi0d}), (\ref{psi0m}) and (\ref{psi0u}). If $Q_m=2l$, one part of the Dirac string is located between two masses, and the other stretches from the lower mass to $-\infty$, and if $Q_m=-2l$, one part of the Dirac string is located between two masses, and the other stretches from the upper mass to $+\infty$.

If we consider the case $\kappa_m=2l$, we have the behaviors of $\psi$:

\noindent
i) $z<-\sigma$ region:
\be
\psi&=&Q_m\label{psi10d}.
\ee
ii) $-\sigma<z<\sigma$ region:
\be
\psi&=&\frac{2l^2-2\sigma^2-\sigma m}{l}\label{psi10m}.
\ee
iii) $z>\sigma$ region:
\be
\psi&=&-Q_m\label{psi10u}.
\ee
This choice in $\psi$ makes no changes in the spacetime structure. In addition to this, if we assume that the total magnetic charge $Q_m$ vanishes, i.e., $Q_m=0$, we then have the solution where the Dirac string does not stretch to spatial infinity but lies between the two magnetically charged masses. In this case, the solutions describe the spacetime created by the magnetic dipole. The dipoles are connected by the Dirac string and the strut. The masses and charges of each pole are given by
\be
&&m_1=\frac{m}{2},\quad Q_{m1}=\frac{lm}{2\sigma}\\
&&m_2=\frac{m}{2},\quad Q_{m2}=-\frac{lm}{2\sigma},
\ee
respectively. 
This configuration contrasts with the 4-soliton Weyl class solution, which also describes the gravitational field created by two magnetically charged masses and a strut. In this 4-soliton solution, at least one of the Dirac strings stretches to $+\infty$ or $-\infty$, and if we consider the case where the Dirac string does not stretch to infinity, there appears a curvature singularity in the solution. In contrast, in this non-Weyl class solution there is no curvature singularity outside the event horizons located at $(\rho,z)=(0,-\sigma)$ and $(0,+\sigma)$ as far as the conditions $l>\sigma$ and $m>-2(\sigma-l)$ hold.

Lastly, we give a short comment on the equilibrium cases in the solutions.
In the 4-soliton solution of the Weyl class, there is an equilibrium case where the strut between two magnetically charged masses disappears. In the solution of the non-Weyl class, the equilibrium condition $\sigma=l$ leads to the solution where one of the two poles disappears. The solution falls into the Weyl class and becomes the extremal Reissner-Nordstr\" om black hole solution with magnetic charge and the Dirac string. In this case, the Dirac string stretching to $-\infty$ or $+\infty$ re-emerges on the $z$-axis.

%%%%%%%%%%%%%%%%%%%%%%%%%%%%%%%%%%%%%%%%%%%%%%%%%%%%%%%%%%%%%%%%%%%%%
\section{Summary and Discussion}
%%%%%%%%%%%%%%%%%%%%%%%%%%%%%%%%%%%%%%%%%%%%%%%%%%%%%%%%%%%%%%%%%%%%%
In this paper we studied the axisymmetric and static Einstein-Maxwell equations.
One of the well-known solutions is the Reissner-Nordstr\"om solution. In deriving this solution,  a functional relationship between the metric component and the electromagnetic potentials is assumed. Such solutions are referred to as the Weyl class (Weyl's electrovac or magnetovac) solutions. When this assumption is not made, the equations become slightly more difficult to solve, and the resulting solutions are called non-Weyl class solutions. We have examined non-Weyl class solutions in detail in this paper.

To obtain non-Weyl class solutions of the Einstein-Maxwell equations, we applied the gauge-theoretical method introduced in our previous work. We considered the matrix function $h(\rho,z)$ composed of the metric function $f(\rho,z)$ and the function $\tilde\chi(\rho,z)$ which is related to the electromagnetic potentials (see Eqs.(\ref{tildechi}) and (\ref{matrixh})). We showed that $h(\rho,z)$ satisfies one of the equations derived from the zero-curvature condition and provided soliton solutions for $h(\rho,z)$.
We demonstrated that the 2-soliton solution in the electrostatic case coincides with Bonnor's solution, and that the 2-soliton solution in the magnetostatic case describes the gravitational field generated by two magnetically charged masses and the Dirac strings.

In the metric (\ref{metric}), Weyl class solutions of the Einstein-Maxwell equations including the Reissner-Nordstr\"om solution satisfy the relations
\be
f=1-2c\chi+\chi^2,\quad {\rm or}\quad f=1+2c_m\tilde\chi_m+\tilde\chi_m^2.\label{fchi2}
\ee
Here $\chi$ is the electrostatic potential and $\tilde\chi_m$ is a function related to the magnetostatic potential $\psi$ through $\tilde\chi_m,_\rho=-\rho^{-1}f\psi,_z, \tilde\chi_m,_z=\rho^{-1}f\psi,_\rho$. The constants $c$ and $c_m$ are related to the total mass $m$ and total electric charge $e$ or magnetic charge $Q_m$ by
\be
c=\frac{m}{e},\quad {\rm or}\quad c_m=\frac{m}{Q_m}.\label{cme}
\ee
On the other hand, such relations do not exist in non-Weyl class solutions, as can be seen  from their asymptotic behaviors in Eqs. (\ref{finf}), (\ref{chiinf}), (\ref{fminf}) and (\ref{psiinf}). 

We now discuss the similarities and contrasts between the Weyl class and the non-Weyl class solutions in the magnetostatic case. In both cases, 1-soliton solutions do not satisfy the asymptotically flat condition. The 2-soliton solution of the Weyl class describes a magnetically charged monopole black hole with a Dirac string. The 4-soliton solution describes the gravitational field generated by two magnetically charged masses with a strut between them and Dirac strings. In contrast, the 2-soliton solution in the non-Weyl class has a structure similar to the 4-soliton solution in the Weyl class. Their sources are located inside regions that possess the nature of an event horizon. There are no curvature singularities outside these regions,  although conical singularities---referred to as struts---exist between the sources. 

Dirac strings appear in both solutions. There are two cases: i) They extend to spatial infinity along the symmetry axis. ii) One string extends to infinity, while the other lies between the sources. In addition, the non-Weyl class solution allows  a configuration in which the Dirac string lies between the sources. In this case, the total magnetic charge vanishes and the individual charges have opposite signs. This configuration does not occur in the Weyl class solution because opposite charges would imply a negative individual mass due to the relation (\ref{cme}) (see Ref.\cite{AK0}), which leads to a curvature singularity. 

There is an equilibrium case in the Weyl class solution where the strut between the sources disappears. In this case, the solution becomes of the Majumdar-Papapetrou type\cite{maj, papa}. In contrast, the non-Weyl class solution does not admit such an equilibrium configuration. Instead, the condition for equilibrium causes one of the two sources to vanish, reducing the solution to the extremal Reissner-Nordstr\"om solution with magnetic charge and a Dirac string. This is a notable feature, as the equilibrium condition transforms a two-source non-Weyl class solution into a single-source solution of the Weyl class.
%%%%%%%%%%%%%%%%%%%%%%%%%%%%%%%%%%%%%%%%%%%%%%%%%%%%%%%%%%%%%%%%%%%%%%%%%%%%%%%


\begin{thebibliography}{9}
\bibitem{AK}T. Azuma and T. Koikawa, Prog. Theor. Exp. Phys.  {\bf 2025}, 023B06 (2025).
\bibitem{miura}R.M. Miura, J. Math. Phys. {\bf 9}, 1202 (1968).
\bibitem{GGKM}C.S. Gardner, J.M. Green, M.D. Kruskal and R.M. Miura, Phys. Rev. Lett. {\bf 19}, 1095 (1967) 
\bibitem{ZakShab1}V.E. Zakharov and A.B. Shabat, Sov. Phys. {\bf JETP 34}, 62 (1972).
\bibitem{ZakShab2}V.E. Zakharov and A.B. Shabat, Sov. Phys. {\bf JETP 37}, 823 (1973).
\bibitem{Wadachi1}M. Wadachi, J. Phys. Soc. Jaoab {\bf 32}, 1681 (1972).
\bibitem{Wadachi2}M. Wadachi, J. Phys. Soc. Jaoab {\bf 34}, 1289 (1973).
\bibitem{akns}M.J. Ablowitz, D.J. Kaup, A.C. Newell and H. Segur, Phys. Rev. Lett. {\bf 31}, 125 (1973).
\bibitem{aks}J.M. Alberty, T. Koikawa and R. Sasaki, Physica {\bf D5}, 43 (1982).
\bibitem{BZ1}V.A. Belinski and V.E. Zakharov, Sov. Phys. {\bf JETP 48}, 985 (1978).
\bibitem{BZ2}V.A. Belinski and V.E. Zakharov, Sov. Phys. {\bf JETP 50}, 1 (1979).
\bibitem{BV}V. Belinski and E. Verdaguer, {\it Gravitational Solitons.} Cambridge University Press,  2001.
\bibitem{B1}W.B. Bonnor, J. Phys. A: Math. Gen. {\bf 12}, 853 (1979).
\bibitem{reiss}H. Reissner, Annalen der Physik {\bf 50}, 106 (1916).
\bibitem{nord}G. Nordstr\"om, Proc, Kon. Ned. Akad. Wet. {\bf 20} 1238 (1918).
\bibitem{BS}W.B. Bonnor and B.R. Steadman, Gen. Rel. Grav. {\bf 43}, 1777 (2011).
\bibitem{B2}W.B. Bonnor, Gen. Rel. Grav. {\bf 44}, 3009 (2012).
\bibitem{AK0}T. Azuma and T. Koikawa, Prog. Theor. Phys.  {\bf 92}, 1095 (1994).
\bibitem{JC}J. Carminati, Gen. Rel. Grav. {\bf 13}, 1185 (1981).
\bibitem{CMR}I. Cabrera-Munguia, V.S. Manko and E. Ruiz, Gen. Rel. Grav. {\bf 43}, 1593 (2011).

\bibitem{maj}S.D. Majumdar, Phys. Rev. \textbf{72}, 390 (1947).
\bibitem{papa}A. Papapetrou, Proc. Roy. Irish Acad. \textbf{A51}, 191 (1947).
\end{thebibliography}
\end{document}